\documentclass[11pt]{article}

\usepackage{amssymb,amsmath,amsfonts}
\usepackage{latexsym}
\usepackage{geometry}
\usepackage[unicode=true]{hyperref}
\usepackage{graphicx}
\usepackage{subcaption}
\usepackage{caption}

\newcommand{\be}{\begin{eqnarray}}
\newcommand{\ee}{\end{eqnarray}}

\begin{document}

\title{Building Momentum Kernel from Shapovalov Form}
\author{Chih-Hao Fu${}^{a,b}$ and Yihong Wang${}^{a,c}$ 
\vspace{0.5cm} \\ 
\it{${}^{a}$School of Fundamental Physics and Mathematical Sciences}\\  
\it{Hangzhou Institute for Advanced Study, UCAS, Hangzhou 310024, China.}\\
\it{${}^{b}$School of Physics and Information Technology,
  Shaanxi Normal University}\\  
\it{No.620 West Chang'an Avenue, Xi'an 710119, P.R. China. }\\
\it{${}^{c}$Laboratoire d'Annecy-le-Vieux de Physique Th\`eorique (LAPTh),}\\  
\it{CNRS and université Savoie Mont-Blanc, }\\ 
  \it{Annecy, 74940, France}
}
\date{  }
\maketitle

\begin{abstract} 
These notes are an extended version of the talks given by the authors
at the XIV International Workshop on Lie Theory and Its Applications
in Physics, Sofia, Bulgaria, 20-26 June 2021. The concise version
published in the proceedings of the workshop contains additional discussions
for the $q$-deformed scenario:

\noindent\href{https://link.springer.com/chapter/10.1007/978-981-19-4751-3_23}{https://link.springer.com/chapter/10.1007/978-981-19-4751-3\_23}.

In these notes we identify KLT kernel with the Shapovalov form on
Verma module with its highest/lowest weight given by the reference momentum
and rest of the momenta as roots. We then take a step forward and
show how the Feynman diagrams emerge naturally as the Shapovalov duals
of the Verma module basis vectors. We show such algebraic construct
offers a compact expression for the BCJ numerators. Explicit examples
are shown for the nonlinear sigma model and the HEFT pre-numerators.
\end{abstract}

\section{Root Systems, Shapovalov Forms, and their implications in BCJ Duality}

\label{sec:klt-kernel}

Recall that in the usual definition a semisimple Lie algebra $g$
is classified by its root system, which is a collection of $D$-dimensional
vectors $\{k_{i}\},i=2,...$, and the algebra $g$ itself is constructed
from the Cartan-Weyl basis generators $E_{k_{i}}$, $F_{k_{i}}$ labelled
by its roots, and its Cartan subalgebra $H_{\mu}$, satisfying the
following relations.
\begin{equation}
[E_{k_{i}},F_{k_{j}}]=\delta_{ij}k_{i}^{\mu}H_{\mu},\hspace{0.5cm}[H_{\mu},E_{k_{j}}]=(k_{j})_{\mu}E_{k_{j}},\hspace{0.5cm}[H_{\mu},F_{k_{j}}]=-(k_{j})_{\mu}F_{k_{j}},\hspace{0.5cm}\mu=1,...,D.\label{eq:cartan-basis-relations}
\end{equation}

In standard QCD calculations it is often more convenient to introduce
gauge group dependence directly through its matrix representation
instead of through its root system, as it is physically difficult
to measure the specific colour of each particle. In the following
discussions however we will retain explicit root labels. As we will
soon see, fundamental physical objects in the context of BCJ duality
have compact expressions in a basis where root labels are manifest.
For our purpose it is actually more convenient to work in the Chevalley
basis, where elements of the Cartan subalgebra appear as scalar products
$H_{k_{i}}=k_{i}^{\mu}H_{\mu}$. We will also denote Lie algebra generators
$E_{k_{i}}$ and $F_{k_{i}}$ as $E_{i}$ and $F_{i}$ respectively
for simplicity. In this new basis the Lie algebra is given by
\begin{equation}
[E_{i},F_{j}]=\delta_{ij}H_{k_{i}},\hspace{0.5cm}[H_{k_{i}},E_{j}]=(k_{i}\cdot k_{j})E_{j},\hspace{0.5cm}[H_{k_{i}},F_{j}]=-(k_{i}\cdot k_{j})F_{j}.\label{eq:chevalley-basis-relations}
\end{equation}

Additionally, we need a representation for this Lie algebra $g$. For
generic Lie algebra it is customary to introduce a Verma module $M_{k_{1}}$,
as opposed to the case of angular momentum algebra where it is more
common to introduce a linear vector space labelled by eigenvalues.
The reason being that beyond $sl_{2}$ we have more pairs of raising
and lowering operators $\{E_{i},\,F_{i}\}$ and it is necessary that
we distinguish the order they act on vectors. The Verma module is
defined using a single lowest (highest) weight vector $V_{1}$, corresponding
in the case of $sl_{2}$ to the vector labelled by lowest (highest)
eigenvalue. The vector carries a lowest weight $k_{1}$ in the sense
that it satisfies the following conditions
\begin{equation}
F_{i}V_{1}=0,\hspace{0.5cm}H_{k_{i}}V_{1}=\left(k_{1}\cdot k_{i}\right)V_{1},\label{eq:verma-module}
\end{equation}
for all lowering operators $F_{i}$ and Cartan subalgebra operators
$H_{k_{i}}$, $i=2,..$, while rest of the vectors in this representation
$M_{k_{1}}$ are polynomials of raising operators acting on this unique
vector $E_{2}V_{1}$, $E_{2}E_{3}V_{1}$, $(E_{2})^{2}V_{1}$, $(E_{2})^{3}V_{1}$,
$\dots$ and so on.

In these notes, we are interested in algebra expressions of the duality
structure discovered by Bern, Carrasco and Johansson (BCJ) \cite{Bern:2008qj,Bern:2010ue}.
This is a highly efficient formulation of the scattering amplitudes,
known to apply to a wide variety of theories including QCD and gravity.
In particular, the double copy formulation offers a systematic method
for generating nonlinear classical gravity solutions from gauge theory
 and state-of-the-art post-Minkowskian
expansion calculations to the gravitational two-body problem,
therefore it is of great importance in modern scattering amplitudes
research \cite{Kosower:2022yvp,Bjerrum-Bohr:2022blt}. One main feature of the BCJ double copy is that it provides
a unified picture of the colour gauge group and the momentum dependence
of the amplitudes, where they are both encapsulated in the form of
BCJ numerators, so that the amplitude of a particular theory of interests
formally becomes a bi-adjoint $\phi^{3}$ amplitude, with colour and/or
kinematic numerators behaving as the two gauge groups of the bi-adjoint
theory. See \cite{Bern:2022wqg} for a recent review on the subject.

To incorporate such a setting, note that albeit its assumed algebra
structure, the kinematic numerator presents itself in the amplitude
as a $\mathbb{C}$-number (as opposed to an algebra or matrix-value
function) so that it is natural to suspect that the algebra dependence
is introduced through a dual map: $g\rightarrow\mathbb{C}$. As it
turns out it does not cost much to extend one step further and assume
the dependence through a bilinear form (bracket), since in that case
we may reduce the notion of a bilinear form to a dual map by feeding
one of its two slots with the unity or a fixed algebra element. Indeed
the bilinear form construction was pioneered by Frost, Mafra, and Mason in \cite{Frost:2020eoa}.
From this perspective the colour dependence of the scattering amplitude
is also introduced through a bilinear form $tr(T^{a}\,T^{b})$, or
Killing form in the case of adjoint representation.

In reality, the choices for a bilinear form are actually quite limited
if we demand the commonly assumed well-behaved properties such as
being a symmetric form and Lie algebra invariance. For the numerous
reasons cited above let us consider the Shapovalov form $\left\langle \hspace{0.2cm},\hspace{0.2cm}\right\rangle $,
which is the symmetric bilinear form defined on the Verma module $M_{k_{1}}$
satisfying additionally
\begin{equation}
\left\langle E_{i}V,U\right\rangle =\left\langle V,F_{i}U\right\rangle \label{eq:shapovalov-form-defn}
\end{equation}
for $U$, $V$ any members in the Verma module (polynomial actions
of the raising operators on $V_{1}$). The Shapovalov form is normalised
so that $\left\langle V_{1},V_{1}\right\rangle =1$ and yields zero
if two slots contain vectors of different weights. This definition
is unique, in the sense that one can start from the normalisation
condition on $V_{1}$ and derive iteratively values of all possible
pairs using (\ref{eq:shapovalov-form-defn}).

The value of a Shapovalov form can be readily calculated using the
common textbook procedure in Quantum Mechanics, namely we translate
all the $E_{i}$'s on one side into the actions of $F_{i}$'s on the
other one by one using the defining property (\ref{eq:shapovalov-form-defn})
of Shapovalov form, and then move each lowering operator $F_{i}$
passing all existing operators using the algebra relations (\ref{eq:chevalley-basis-relations})
until the lowering operator finally acts on the lowest weight vector
$V_{1}$ and vanishes because of (\ref{eq:verma-module}). Explicitly,
the Shapovalov form of the lowest weight vector $V_{1}$ is defined
to be unity. The next lowest weight vector is $E_{i}V_{1}$, translating
using (\ref{eq:shapovalov-form-defn}) yields
\begin{align}
\left\langle E_{i}V_{1},E_{i}V_{1}\right\rangle  & =\left\langle V_{1},[F_{i},E_{i}]V_{1}\right\rangle =\left\langle V_{1},H_{k_{i}}V_{1}\right\rangle \\
 & =k_{1}\cdot k_{i}\nonumber 
\end{align}
These results can be simplified if we introduce shorthand notations
for scalar products of two roots $s_{ij}=k_{i}\cdot k_{j}$. Also
for later discussions, we denote $s_{A}=\sum_{i,j}s_{ij}$ for a finite
set of roots $A\subset\{1,...,n\}$. The next-to-next-to-lowest weight
vectors consist of pairs of vectors carrying degenerate weights. Let
us for the moment focus on the pair $E_{3}E_{2}V_{1}$ and $E_{2}E_{3}V_{1}$.
These two share the same weight $k_{1}+k_{2}+k_{3}$ and can have
non-vanishing Shapovalov form. Proceed as above shows that the Shapovalov
form of these two vectors are
\begin{align}
\left\langle E_{3}E_{2}V_{1},E_{3}E_{2}V_{1}\right\rangle  & =(s_{23}+s_{13})s_{12},\\
\left\langle E_{3}E_{2}V_{1},E_{2}E_{3}V_{1}\right\rangle  & =s_{13}s_{12}.\nonumber 
\end{align}
An amplitude theorist would immediately recognise the above results
are in fact the momentum kernels that appear in the KLT relations
\cite{Kawai:1985xq} of a four graviton scattering process, if we identify
the lowest weight $k_{1}$ and roots $k_{2}$, $k_{3}$ as momenta
of leg $1$, $2$ and $3$ respectively. These are the celebrated
relations that identify gravity amplitudes as the squares of gauge
theory ones, even though the two theories are not apparently related
at the field theory Lagrangian level. These relations are better understood
in string theory, and a systematic formula for momentum kernel is
attainable by taking the field theory limit \cite{Bjerrum-Bohr:2010pnr,Broedel:2013tta}.
For an $n$-point scattering we have
\begin{equation}
\mathcal{S}[i_{2},\dots,i_{n-1}|j_{2},\dots,j_{n-1}]=\prod_{t=2}^{n-1}(k_{1}\cdot k_{i_{t}}+\sum_{q>t}^{n-1}\theta(i_{t},i_{q})k_{i_{t}}\cdot k_{i_{q}}),\label{eq:momentum-kernel}
\end{equation}
where $\{i_{2},\dots,i_{n-1}\}$ and $\{j_{2},\dots,j_{n-1}\}$ denote
the order of legs $2$ to $n-1$ that appear in the two copies of
gauge theory amplitudes and $\theta(i_{t},i_{q})=1$ if the relative
order of $i_{t}$ and $i_{q}$ are opposite in these two sets and
yields zero otherwise.

The equivalence between the Shapovalov form and momentum kernel continues
as we go through higher weight vectors. For our purposes, it is actually
more convenient to denote basis vectors in the Verma module using
their weights and the order that raising operators act. At $4$-point
we write the pair of vectors sharing the same weight $k_{1}+k_{2}+k_{3}$
as the following.
\begin{align}
V(k_{1}+k_{2}+k_{3},I) & =E_{3}E_{2}V_{1},\\
V(k_{1}+k_{2}+k_{3},(23)) & =E_{2}E_{3}V_{1},\nonumber 
\end{align}
where we used permutation maps $I$ and $(23)$ to label the two respective
orders of $E_{2}$ and $E_{3}$. Following the same spirit, an $n$-point
scattering process corresponds to the sector that has weight $\sum_{i=1}^{n-1}k_{i}$,
and the basis vectors in these sectors are given by
\begin{equation}
V(\sum_{i=1}^{n-1}k_{i},\sigma)=E_{\sigma(n-1)}...E_{\sigma(2)}V_{1},\quad\sigma\in S_{n-2}.\label{eq:basis}
\end{equation}
The equivalence between the Shapovalov form and momentum kernel can be
more explicitly written as the following.
\begin{equation}
V(\sum_{i=1}^{n-1}k_{i},\tau),V(\sum_{i=1}^{n-1}k_{i},\sigma)\rangle=\mathcal{S}[\tau(n-1),\dots,\tau(2)|\sigma(2),\dots,\sigma(n-1)].\label{eq:shapovalov-klt-equiv}
\end{equation}

We would like to emphasise that the discussions in these notes are
independent of the physical origins of the kinematic algebra. The
purpose of these discussions is to explore how much can be derived,
assuming only the existence of a semi-simple Lie algebra structure.
The explicit algebra structure can be furnished either by the diffeomorphism
algebra explained in \cite{Monteiro:2011pc,Bjerrum-Bohr:2012kaa},
the fusion rules algebra of \cite{Chen:2019ywi,Chen:2021chy},
or the vertex algebra discussed in \cite{Fu:2018hpu,Fu:2020frx}
for string theory settings\footnote{One caveat: The presumed momenta in an explicit construction may not
necessarily coincide with their root system even if they form a semi-simple
Lie algebra. At the moment it is not clear to us whether extra conditions
are required in these physical representations. In the following discussions
the term momentum is exclusively reserved for the roots of the kinematic
algebra unless specified otherwise.}. We will see in section \ref{sec:amplitude-numerator-formulas} that
both amplitudes and numerators can be expressed in a rather unified
formulation in terms of Shapovalov forms, where all the theory-dependent
information is encoded as a fixed reference vector of the Verma module.

\section{Dual basis induced by Shapovalov form}

\label{sec:dual-basis}

In a KLT relation describing $n$ physical particles scattering, the
corresponding momentum kernel (\ref{eq:momentum-kernel}) is written
entirely in terms of the momenta of the first $n-1$ particles. Its
dependence on the last leg $k_{n}$ is only implicitly introduced
through momentum conservation. In its matrix form, indexed by all
possible $(n-2)!$ permutations of legs from $2$ to $n-1$, the momentum
kernel is known to be invertible for generic momentum configuration
and when the resulting $k_{n}$ being not on the null shell (we may
need to analytic continue $k_{n}$ if necessary) \cite{Kliermaier:2010,Du:2011js,Cachazo:2013iea,Mafra:2016ltu}.
Under these circumstances the Shapovalov form as a whole defined on
the full Verma module is non-singular because it is identified with
the momentum kernel within each weight $\sum_{i=1}^{n-1}k_{i}$ subspace
(\ref{eq:shapovalov-klt-equiv}). Suppose if we then define dual vectors
by orthogonality conditions,
\begin{equation}
\left\langle V^{*}(\sum_{i=1}^{\ell}k_{i},\sigma),V(\sum_{i=1}^{\ell}k_{i},\tau)\right\rangle =\delta_{\sigma,\tau},\label{eq:dualdef}
\end{equation}
for all particle numbers $n:=\ell+1$ ($\ell$ being the length of
word consists of raising operators and $V_{1}$), weight subspaces,
and for all permutations, the collection of all such dual vectors
$V^{*}(\sum_{i=1}^{\ell}k_{i},\sigma)$ forms a basis of the same
linear vector space $M_{k_{1}}$. The orthogonality condition can
be made manifest if we decorate a vector with star sign, such as $(E_{\sigma(\ell)}\dots E_{\sigma(3)}E_{\sigma(2)}V_{1})^{*}$,
to represent the dual vector that is orthogonal to all permutations
except $E_{\sigma(\ell)}\dots E_{\sigma(3)}E_{\sigma(2)}V_{1}$. In
the following discussions we will be using these two equivalent notations
interchangeably, depending on which better suits our purpose.
\begin{equation}
(E_{\sigma(\ell)}\dots E_{\sigma(3)}E_{\sigma(2)}V_{1})^{*}=V^{*}(\sum_{i=1}^{\ell}k_{i},\sigma).
\end{equation}

The basis vectors in the dual basis just defined satisfy a recursion
relation between vectors formed by words of different lengths.
\begin{equation}
F_{j}(E_{\sigma(\ell)}\dots E_{\sigma(3)}E_{\sigma(2)}V_{1})^{*}=\delta_{j,\,\sigma(\ell)}(E_{\sigma(\ell-1)}\dots E_{\sigma(3)}E_{\sigma(2)}V_{1})^{*}.\label{eq:recuriosn-dual-vectors}
\end{equation}
The fact that it is true can be easily seen from the iterative property
of Shapovalov form (\ref{eq:shapovalov-form-defn}). Because of the
action of a lowering operator $F_{j}$, the left hand side of (\ref{eq:recuriosn-dual-vectors})
is now a vector in the weight subspace $(\sum_{i=1}^{\ell}k_{i})-k_{j}$.
To yield a nonvanishing Shapovalov form of this vector, the other
slot must carry the same weight. For generic momentum configuration
this is therefore spanned by the set of vectors obtained from permutation
actions of $\{E_{2},\,E_{3},\,\dots,\,E_{\ell}\}$, but without $E_{j}$.
That is, the set of vectors of the form $E_{\tau(\ell)}...E_{\tau(j+1)}E_{\tau(j-1)}...E_{\tau(2)}V_{1}$,
for all possible permutations $\tau$ of the $\ell-1$ labels. Moving
$F_{j}$ to the right using (\ref{eq:shapovalov-form-defn}) gives
\begin{equation}
\left\langle (E_{\sigma(\ell)}E_{\sigma(\ell-1)}\dots E_{\sigma(3)}E_{\sigma(2)}V_{1})^{*},E_{j}E_{\tau(\ell)}...E_{\tau(j+1)}E_{\tau(j-1)}...E_{\tau(2)}V_{1}\right\rangle .
\end{equation}
The Shapovalov form is only non-zero when $\sigma(\ell)=j$ and the
rest of the labels $\{\sigma(\ell-1),\dots,\sigma(2)\}$ agree with
$\{\tau(\ell-1),\,\dots,\,\tau(j+1),\,\tau(j-1),\,\dots,\tau(2)\}$
both in order and values, because of the definition of a dual vector
(\ref{eq:dualdef}), and this proves the recursion relation (\ref{eq:recuriosn-dual-vectors}).

The converse is actually also true. Starting with $(V_{1})^{*}=V_{1}$,
suppose we construct iteratively a set of vectors by the recursion
relation (\ref{eq:recuriosn-dual-vectors}), using vectors with words
of length $\ell$ to generate those with length $\ell+1$, the resulting
set of vectors is the same dual basis vectors that satisfy the defining
orthogonality conditions (\ref{eq:dualdef}). The proof can be seen
by noticing that the original dual vectors $V^{*}(\sum_{i=1}^{\ell}k_{i},\sigma)$
form a basis, so that if we denote the iteratively constructed vector
as $U$, it should be expressible as a linear combination
\begin{equation}
U=\sum_{\sigma\in S_{\ell-1}}C_{\sigma}\,V^{*}(\sum_{i=1}^{\ell}k_{i},\sigma).
\end{equation}
Plugging the above into Shapovalov form and moving the $F_{j}$ in
the definition of $U$ to the other side proves that the coefficient
$C_{\sigma}=1$ for the term labelled by exactly the same permutation,
and zero otherwise.

The set of basis vectors $V(\sum_{i=1}^{\ell}k_{i},\sigma)$ and their
duals $V^{*}(\sum_{i=1}^{\ell}k_{i},\sigma)$ together work as brackets
in the familiar way. Transformations between the two can be easily
seen from the orthogonality condition to be the following.
\begin{align}
V(\sum_{i=1}^{\ell}k_{i},\sigma)= & \left\langle V(\sum_{i=1}^{\ell}k_{i},\sigma),V(\sum_{i=1}^{\ell}k_{i},\tau)\right\rangle V^{*}(\sum_{i=1}^{\ell}k_{i},\tau),\label{eq:dtob}\\
V^{*}(\sum_{i=1}^{\ell}k_{i},\sigma)= & \left\langle V^{*}(\sum_{i=1}^{\ell}k_{i},\sigma),V^{*}(\sum_{i=1}^{\ell}k_{i},\tau)\right\rangle V(\sum_{i=1}^{\ell}k_{i},\tau).\label{eq:btod}
\end{align}

\section{Propagator matrix and inverse Shapovalov form}

\label{sec:inverse-klt-proof}

The transformations between basis vectors and their duals immediately
imply that the Shapovalov form of the basis vectors $\left\langle V(\sum_{i=1}^{\ell}k_{i},\sigma),V(\sum_{i=1}^{\ell}k_{i},\tau)\right\rangle $
and the Shapovalov form of the dual basis vectors $\left\langle V^{*}(\sum_{i=1}^{\ell}k_{i},\sigma),V^{*}(\sum_{i=1}^{\ell}k_{i},\tau)\right\rangle $
are in fact inverse to each other. (To see this simply plug (\ref{eq:btod})
into (\ref{eq:dtob}).) We know from earlier arguments that the Shapovalov
form of basis vectors is the momentum kernel that appears in the KLT
relations. Its inverse is known by amplitude theorists to be the current
of colour-stripped bi-adjoint $\phi^{3}$ amplitude $J(\sigma|\tau)$,
\begin{equation}
\left\langle V^{*}(\sum_{i=1}^{l}k_{i},\sigma),V^{*}(\sum_{i=1}^{l}k_{i},\tau)\right\rangle =J(\sigma|\tau).\label{eq:sfm}
\end{equation}
In the following we will sometimes loosely refer $J(\sigma|\tau)$
as the propagator matrix, even though strictly speaking the two are
related by a rescaling, $J(\sigma|\tau)=\frac{1}{k_{n}^{2}}m(\sigma|\tau)$.
Incidentally, the bi-adjoint $\phi^{3}$ current can be understood
as binary trees\footnote{The notation we use here is related to the binary tree in \cite{Frost:2020eoa,Mafra:2020qst}
by $J(\sigma|123\dots n)=b(\sigma)$. }. For a very readable discussion of the algebraic combinatorics approach
of the propagator matrix, see \cite{Mafra:2020qst}. Because of
its importance, the fact that the momentum kernel is inverse to the bi-adjoint
current has been proved in the physics literature from various perspectives
\cite{Kliermaier:2010,Du:2011js,Cachazo:2013iea,Mafra:2016ltu}.
In the remaining part of this section, we give a very sketchy outline
of the proof in terms of algebra language, only for the purpose of
completeness. Readers already familiar with this content may wish
to jump directly to section \ref{sec:amplitude-numerator-formulas}
for Shapovalov form formulations of the amplitude and numerator.

To prove that the inverse momentum kernel is indeed the bi-adjoint
current, first note that the off-shell continued BCJ relation can
be readily derived from transforming the recursion relation (\ref{eq:recuriosn-dual-vectors})
of dual basis vectors into the original basis vectors, and collect
terms. For example at $5$-point, one recursion relation of the dual
basis vectors read
\begin{equation}
F_{4}(E_{4}E_{2}E_{3}V_{1})^{*}=(E_{2}E_{3}V_{1})^{*}.
\end{equation}
When translated into the original basis vectors, the left-hand side
of the equation gives the following.
\begin{align}
 & F_{4}\Bigl(\left\langle (E_{4}E_{2}E_{3}V_{1})^{*},(E_{4}E_{2}E_{3}V_{1})^{*}\right\rangle E_{4}E_{2}E_{3}V_{1}+\left\langle (E_{4}E_{2}E_{3}V_{1})^{*},(E_{2}E_{4}E_{3}V_{1})^{*}\right\rangle E_{2}E_{4}E_{3}V_{1}\\
 & +\left\langle (E_{4}E_{2}E_{3}V_{1})^{*},(E_{2}E_{3}E_{4}V_{1})^{*}\right\rangle E_{2}E_{3}E_{4}V_{1}+\dots\Bigr)\nonumber \\
 & =(s_{14}+s_{24}+s_{34})\left\langle (E_{4}E_{2}E_{3}V_{1})^{*},(E_{4}E_{2}E_{3}V_{1})^{*}\right\rangle E_{2}E_{3}V_{1}+(s_{14}+s_{24})\left\langle (E_{4}E_{2}E_{3}V_{1})^{*},(E_{2}E_{4}E_{3}V_{1})^{*}\right\rangle E_{2}E_{3}V_{1}\nonumber \\
 & +s_{14}\left\langle (E_{4}E_{2}E_{3}V_{1})^{*},(E_{2}E_{3}E_{4}V_{1})^{*}\right\rangle E_{2}E_{3}V_{1}+\dots.\nonumber 
\end{align}
The coefficients of basis vector $E_{3}E_{2}V_{1}$ from both sides
of the equation therefore together give the identity
\begin{align*}
\left\langle (E_{2}E_{3}V_{1})^{*},(E_{2}E_{3}V_{1})^{*}\right\rangle = & (s_{14}+s_{24}+s_{34})\left\langle (E_{4}E_{2}E_{3}V_{1})^{*},(E_{4}E_{2}E_{3}V_{1})^{*}\right\rangle \\
 & +(s_{14}+s_{24})\left\langle (E_{4}E_{2}E_{3}V_{1})^{*},(E_{2}E_{4}E_{3}V_{1})^{*}\right\rangle \\
 & +s_{14}\left\langle (E_{4}E_{2}E_{3}V_{1})^{*},(E_{2}E_{3}E_{4}V_{1})^{*}\right\rangle .
\end{align*}
Likewise, for generic $n$-point, we have
\begin{align}
 & \sum_{p=1}^{\ell}\left(k_{j}\cdot k_{1}+\sum_{i=2}^{p}k_{j}\cdot k_{\sigma(i)}\right)\left\langle \left(E_{\sigma(\ell)}E_{\sigma(\ell-1)}...E_{\sigma(2)}V_{1}\right)^{*},\left(E_{\tau(\ell)}\dots E_{\tau(p+1)}E_{j}E_{\tau(p)}...E_{\tau(2)}V_{1}\right)^{*}\right\rangle \nonumber \\
 & =\delta_{j,\sigma(\ell)}\left\langle (E_{\sigma(\ell-1)}...E_{\sigma(2)}V_{1})^{*},(E_{\tau(\ell)}..E_{\tau(2)}V_{1})^{*}\right\rangle .\label{eq:sfbcj}
\end{align}
In addition, the bi-adjoint current, defined as a product of propagators,
satisfies both $U(1)$ decoupling identity and the Berends-Giele recursion
relation \cite{Berends:1988zn}, so that the iterative proof
used in \cite{Du:2011js} translates into algebra language
word-for-word.

\section{Shapovalov form formulation for amplitudes and kinematic numerators}

\label{sec:amplitude-numerator-formulas}

Assuming the identity (\ref{eq:sfm}) between bi-adjoint current and
Shapovalov form of the dual basis vectors proved, we can now write
\begin{align}
V(\sum_{i=1}^{\ell}k_{i},\sigma) & =\mathcal{S}[\sigma|\tau]\,V^{*}(\sum_{i=1}^{\ell}k_{i},\tau),\\
V^{*}(\sum_{i=1}^{\ell}k_{i},\sigma) & =J(\sigma|\tau)\,V(\sum_{i=1}^{\ell}k_{i},\tau).
\end{align}
In the following, we will assume the Shapovalov forms between basis
and between dual basis as known, since they are explicitly identified
with the momentum kernel and propagator matrix respectively. In order
to retain the freedom to translate between the two sets of basis we
will also assume $k_{n}^{2}\neq0$, so that the amplitudes and the
kinematic numerators referred to should be understood as their off-shell
continuations.

Our claim is that for every BCJ satisfying theory, there is a fixed
reference vector $U_{\left\{ 1,\dots,n-1\right\} }$ at each $n$-point
which encodes all the essential information such that all colour-ordered
amplitudes and numerators associated with any cubic graph $\Gamma$
(not just the half ladders) can be described by the following compact
formulas.
\begin{align}
N(\Gamma) & =\frac{1}{2\,s_{123\dots n-1}}\left\langle U_{\left\{ 1,\dots,n-1\right\} },V(\Gamma)\right\rangle ,\label{eq:nsf}\\
A(1\sigma n) & =\left\langle U_{\left\{ 1,\dots,n-1\right\} },V^{*}(\sum_{i=1}^{n-1}k_{i},\sigma)\right\rangle .\label{eq:asf}
\end{align}
For the vector $V(\Gamma)$ that appears in the formulas we have a systematic
set of rules that is independent of the theory to assign to each cubic
graph $\Gamma$ a specific element in the vector space $M_{k_{1}}$.
The resulting collection of vectors automatically respects Jacobi
identities associated with the corresponding graphs. Both amplitudes
and numerators share the same fixed reference vector $U_{\left\{ 1,\dots,n-1\right\} }$.


\begin{figure}
\centering
\begin{subfigure}[a]{0.3\textwidth}
\centering
\includegraphics[width=4cm]{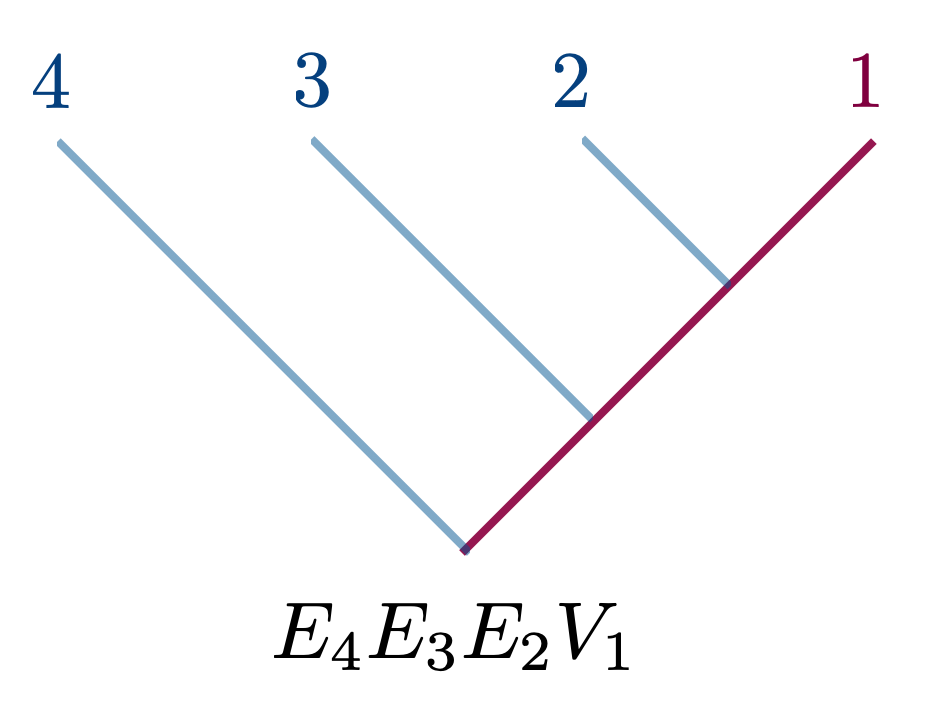}
\caption{  }
\end{subfigure}
\begin{subfigure}[b]{0.3\textwidth}
\centering
\includegraphics[width=4cm]{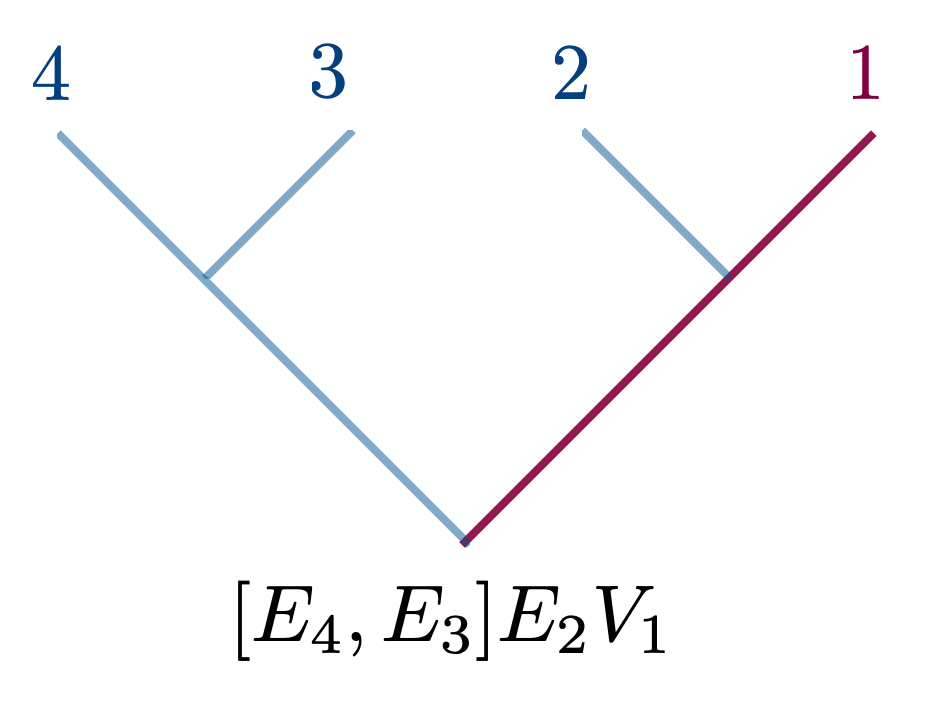}
\caption{  }
\end{subfigure}
\begin{subfigure}[c]{0.3\textwidth}
\centering
\includegraphics[width=4cm]{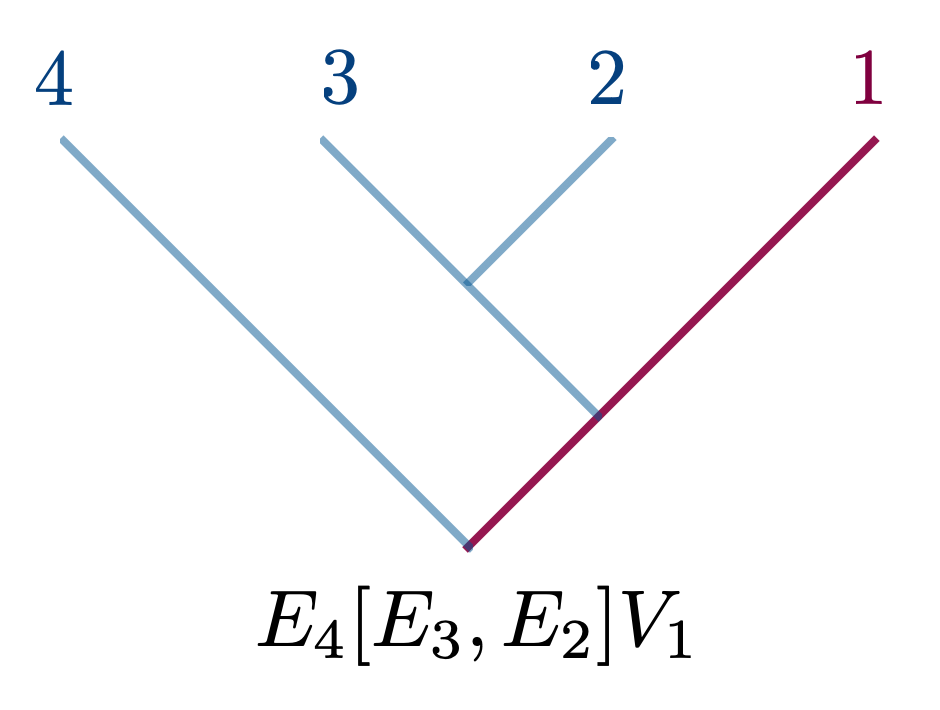}
\caption{  }
\end{subfigure}
\begin{subfigure}[d]{0.3\textwidth}
\centering
\includegraphics[width=4cm]{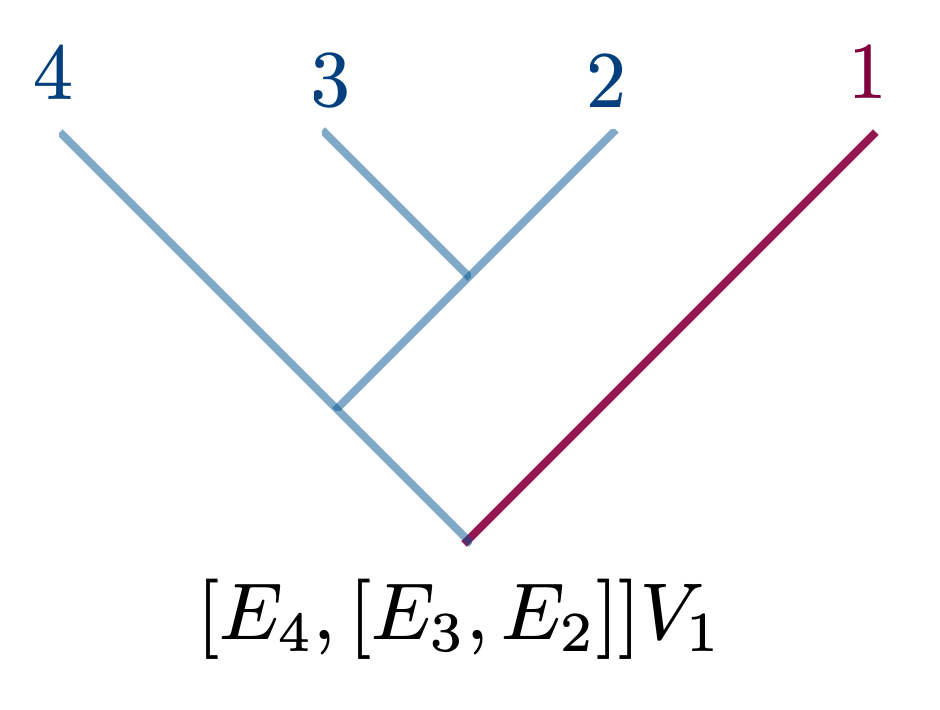}
\caption{  }
\end{subfigure}
\begin{subfigure}[e]{0.3\textwidth}
\centering
\includegraphics[width=4cm]{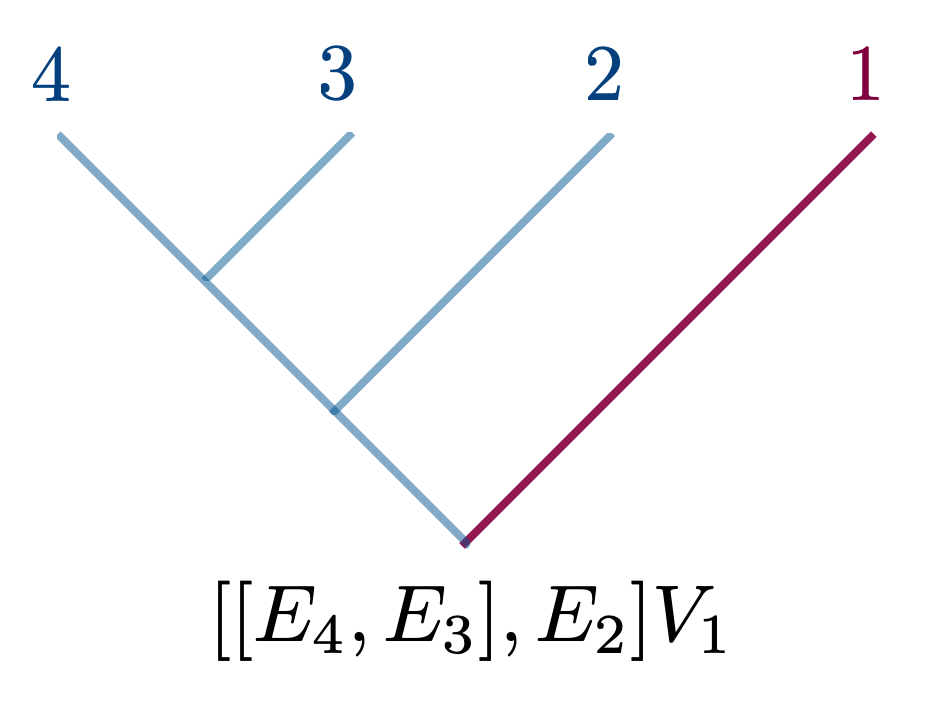}
\caption{  }
\end{subfigure}
\caption{Cubic trees at 5-point. \label{fig:5pt-tree}}
\end{figure}

The rules for assigning graphs to vectors are the following. Let us
consider the cubic graphs at $5$-point as an example (illustrated
in Fig. \ref{fig:5pt-tree} as rooted binary trees with leg $5$ as
their roots). For simplicity, we only draw trees in the amplitude $A(12345)$.
Those in amplitudes of different orders can be obtained by permutation
of the legs/leaves. We mark the path connecting leaf $1$ and the
root red, representing the lowest weight vector $V_{1}$. Rest of
the leaves represent raising operators $E_{i}$, with $i=2$, $3$,
$\dots$, $n-1=4$ in this example. Leaves directly connected to the
marked line are understood as actions on $V_{1}$ in the same order
(for example Fig. \ref{fig:5pt-tree} (a) translates to $E_{4}E_{3}E_{2}V_{1}$).
If this is not the case a leaf $E_{i}$ contributes a commutator $[*,E_{i}]$,
with the $*$ here representing the subtree this leaf is connected
to. (Compare Fig. \ref{fig:5pt-tree} (b) and (c) with Fig. \ref{fig:5pt-tree}
(a).)

A special case for the graphic rules just described is when the structure
constants of the algebra are available, and we consider the adjoint
actions of the Lie algebra $E_{i}=ad(T^{i})$, with $V_{1}$ being
simply the algebra element $T^{1}$. In this case the rules for cubic
graphs simply assign a structure constant to each vertex/node in the
familiar way. Fig. \ref{fig:5pt-tree} (a) in this case is given by
$f_{1,2}{}^{\sigma_{1}}f_{\sigma_{1},3}{}^{\sigma_{2}}f_{\sigma_{2},4}{}^{\sigma_{3}}T^{\sigma_{3}}$.
In either case the Jacobi identities between the assigned vectors
are easily verified\footnote{In the presence of multiple quark lines the colour numerator is more
naturally described by a product of multiple Shapovalov forms, where
the highest/lowest weight vectors and the raising/lowering operators
are given by the vectors and matrices in the fundamental representation.
Incorporating the colour decomposition of \cite{Johansson:2015oia}
into the same unified setting as (\ref{eq:nsf}) where the numerator
is expressed as a single Shapovalov form could be a very interesting
problem from an algebra perspective, which we leave for future work.}.

According to the graphic rules, the vector we assign to each $n$-point
cubic graph has weight $\sum_{i=1}^{n-1}k_{i}$. In order to give
non-vanishing Shapovalov forms, the vector $U_{\left\{ 1,\dots,n-1\right\} }$
that appears in (\ref{eq:nsf}) and (\ref{eq:asf}), if exists, must
also have the same weight, so that we can write this vector as a linear
combination of the basis vectors $V(\sum_{i=1}^{n-1}k_{i},\sigma)$
and try to solve for the coefficients. The fact that these coefficients
are solvable is guaranteed by the nondegeneracy of Shapovalov form.
As a matter of fact the vector $U_{\left\{ 1,\dots,n-1\right\} }$
is directly constructable as the following.
\begin{align}
U_{\left\{ 1,\dots,n-1\right\} } & =\sum_{\sigma\in S_{n-2}}A(1\sigma n)\,V(\sum_{i=1}^{n-1}k_{i},\sigma)\label{eq:uv}\\
 & =\sum_{\sigma\in S_{n-2}}A(1\sigma n)\,E_{\sigma(n-1)}\dots E_{\sigma(2)}V_{1}.\nonumber 
\end{align}
Equation (\ref{eq:asf}) can be easily seen from the orthogonality
of the two basis. To verify (\ref{eq:nsf}) we use the fact that Jacobi
identities are ensured by graphic rules so that we only need to check
half ladders. Denoting the set of half ladder graphs as $\Gamma_{\tau}$,
with $\tau$ the permutation that brings $\{2,3,\dots,n-1\}$ to the
corresponding order, we see that we need to verify the numerators
given by $U_{\left\{ 1,\dots,n-1\right\} }$ satisfies
\begin{equation}
N(\Gamma_{\tau})=\sum_{\sigma\in S_{n-2}}\frac{1}{2\,s_{123\dots n-1}}\mathcal{S}[\sigma|\tau]\,A(1\sigma n).
\end{equation}
Plugging (\ref{eq:uv}) into (\ref{eq:nsf}) and identifies the Shapovalov
form $\left\langle V\left(\sum_{i=1}^{n-1}k_{i},\sigma\right),V\left(\sum_{i=1}^{n-1}k_{i},\tau\right)\right\rangle $
as the momentum kernel then complete the proof. The solution is a
permutation sum so that the reference vector $U_{\left\{ 1,\dots,n-1\right\} }$
is symmetric under permutations of $\{2,3,\dots,n-1\}$, as it is
implicitly assumed by the consistency of (\ref{eq:nsf}) and (\ref{eq:asf}).

In circumstances where explicit numerators are more easily available
(when compared with amplitudes), we may alternatively choose to solve
$U_{\left\{ 1,\dots,n-1\right\} }$ directly from the numerators.
Suppose the half-ladder numerators are known, in terms of dual
basis vectors we have
\begin{equation}
U_{\left\{ 1,\dots,n-1\right\} }=2\,s_{123\dots n-1}\,N(\Gamma_{\sigma})\,V^{*}(\sum_{i=1}^{n-1}k_{i},\sigma).\label{eq:uvn}
\end{equation}
The validity of the above equation can be verified by plugging it into
(\ref{eq:nsf}) and (\ref{eq:asf}).

\section{Algebraic perspective on nonlinear sigma model}

\label{sec:nlsm}

As a first example let us consider the numerators of nonlinear sigma
model. It was discovered in \cite{Chen:2013fya,Chen:2014dfa}
that the nonlinear sigma model is actually a BCJ-satisfying theory
even though it was originally entirely not apparent from the defining
construction. A Lagrangian explanation manifestly presenting the cubic
structure was later constructed by \cite{Cheung:2016prv},
and the BCJ numerators of the nonlinear sigma model were shown to be expressible
in terms of momentum kernels in \cite{Du:2016tbc,Carrasco:2016ldy,Carrasco:2016ygv}.
From the algebra perspective, we argued in section \ref{sec:amplitude-numerator-formulas}
that the reference vector $U_{\left\{ 1,\dots,n-1\right\} }$ appears
in the numerator formula (\ref{eq:nsf}) carries weight $\sum_{i=1}^{n-1}k_{i}$,
so that the reference vector can be spanned by basis vectors in the
same weight subspace,
\begin{equation}
U_{\left\{ 1,\dots,n-1\right\} }=\sum_{\sigma\in S_{n-2}}C_{\sigma}\,V(\sum_{i=1}^{n-1}k_{i},\sigma),
\end{equation}
with $(n-2)!$ coefficients $C_{\sigma}$. Plugging the above into
the numerator formula (\ref{eq:nsf}) we see this in term implies
the numerator can be written as a linear combination of momentum kernels.
Seeing that the numerator formula applies to all theories it is not
a complete surprise that we see this expression for the nonlinear
sigma model. From this perspective it is the fact that the numerator
found by \cite{Carrasco:2016ldy} has a highly compact expression that is indeed
extremely nontrivial.

Let us begin with the formula found by \cite{Carrasco:2016ldy}. In terms of Shapovalov
forms, the $n$-point half ladder basis numerators of the nonlinear sigma
model can be written as the following.
\begin{align}
N(\Gamma_{\sigma}) & =(-1)^{n/2}\pi^{2}\mathcal{S}[\sigma^{T}|\sigma]\label{eq:nlsmlbn}\\
 & =(-1)^{n/2}\pi^{2}\left\langle V(\sum_{i=1}^{n-1}k_{i},\sigma),V(\sum_{i=1}^{n-1}k_{i},\sigma)\right\rangle .\nonumber 
\end{align}
These are not yet in the reference vector expression discussed in
section \ref{sec:amplitude-numerator-formulas}. To proceed, we have
actually two alternatives for the reference vector $U_{\left\{ 1,\dots,n-1\right\} }$.
The nonlinear sigma model amplitudes are known to be vanishing for
odd numbers of particle scattering \cite{Kampf:2012fn,Kampf:2013vha},
and therefore it is the same for kinematic numerators. A natural option
is to pick a vanishing reference vector for odd numbers, so that the
Shapovalov form formulas (\ref{eq:nsf}) and (\ref{eq:asf}) give
the correct numerators and amplitudes for all $n$ particle processes.
Alternatively we may choose a vector $U_{\left\{ 1,\dots,n-1\right\} }$
that yields numerators according to the momentum kernel formula (\ref{eq:nlsmlbn}),
for both even and odd $n$, and identify the result of numerator formula
(\ref{eq:nsf}) with nonlinear sigma model numerators only when $n$
is even. Namely, we may choose the following.
\begin{align}
A(1\sigma n) & =\begin{cases}
\left\langle U_{\left\{ 1,\dots,n-1\right\} },V^{*}(\sum_{i=1}^{n-1}k_{i},\sigma)\right\rangle  & n\in2\mathbb{Z}\\
0 & n\notin2\mathbb{Z}
\end{cases},\label{eq:nlsmasf}\\
N(\Gamma) & =\begin{cases}
\left\langle U_{\left\{ 1,\dots,n-1\right\} },V(\Gamma)\right\rangle  & n\in2\mathbb{Z}\\
0 & n\notin2\mathbb{Z}
\end{cases}.\label{eq:nlsmnsf}
\end{align}
In the discussion below we choose the latter option, as we will see the
nonlinear sigma model numerator in the formulation of \cite{Carrasco:2016ldy}
is simple enough that we can derive a recursion relation for all multiplicity
(provided we extend (\ref{eq:nlsmlbn}) to odd particles).

Seeing that the numerators are more easily available in the case of
nonlinear sigma model, we solve the reference vector $U_{\left\{ 1,\dots,n-1\right\} }$
using (\ref{eq:uvn}), yielding
\begin{equation}
U_{\left\{ 1,\dots,n-1\right\} }=(-1)^{n/2}\pi^{2}2\,s_{123\dots n-1}\sum_{\sigma\in S_{n-2}}\mathcal{S}[\sigma|\sigma]\,V^{*}(\sum_{i=1}^{n-1}k_{i},\sigma)\label{eq:nlsm-ref-vector}
\end{equation}
for both even and odd $n$. Explicitly, at $4$-point the above gives
\begin{align}
U_{\left\{ 1,2,3\right\} }= & 2\,s_{123}\pi^{2}\left\langle E_{3}E_{2}V_{1},E_{3}E_{2}V_{1}\right\rangle (E_{3}E_{2}V_{1})^{*}\\
 & +2\,s_{123}\left\langle E_{2}E_{3}V_{1},E_{2}E_{3}V_{1}\right\rangle (E_{2}E_{3}V_{1})^{*}.\nonumber 
\end{align}
Or, in terms of the basis vectors,
\begin{equation}
U_{\left\{ 1,2,3\right\} }=\pi^{2}\left(s_{12}+s_{23}\right)\text{\ensuremath{E_{3}E_{2}V_{1}}}+\pi^{2}\left(s_{13}+s_{23}\right)E_{2}E_{3}V_{1}.\label{eq:4pt-nlsm-ref-vector}
\end{equation}

\begin{figure}
\centering
\includegraphics[width=4cm]{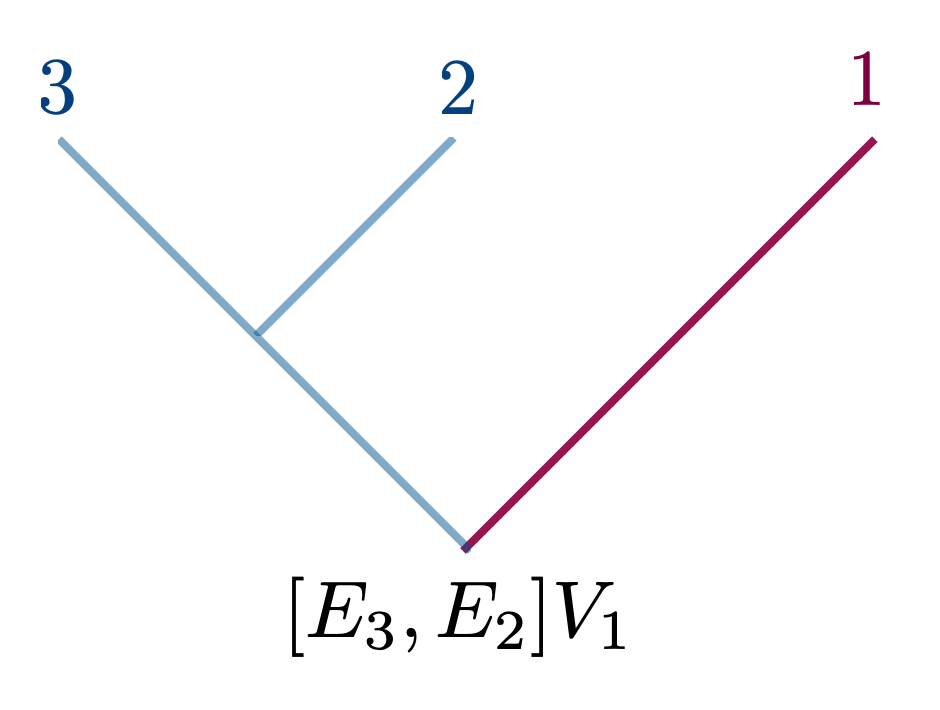}

\caption{The $t$-channel graph at $4$-point. \label{fig:nhl}}
\end{figure}

Let us verify the above result with those produced from the momentum
kernel formula (\ref{eq:nlsmlbn}). At $4$-point we have three channels,
two of them are half ladders and can be written down directly from
(\ref{eq:nlsmlbn}).
\begin{align}
N(\Gamma_{s}) & =\pi^{2}\mathcal{S}[23|23]=\pi^{2}s_{12}\left(s_{13}+s_{23}\right)=\pi^{2}\left\langle E_{3}E_{2}V_{1},E_{3}E_{2}V_{1}\right\rangle ,\\
N(\Gamma_{u}) & =\pi^{2}\mathcal{S}[32|32]=\pi^{2}s_{13}\left(s_{12}+s_{23}\right)=\pi^{2}\left\langle E_{2}E_{3}V_{1},E_{3}E_{3}V_{1}\right\rangle ,\nonumber 
\end{align}
and the remaining one is given by their commutator,
\begin{equation}
N(\Gamma_{t})=N(\Gamma_{s})-N(\Gamma_{u})=\pi^{2}s_{23}\left(s_{12}-s_{13}\right).
\end{equation}
We have an immediate agreement for the two half ladders if we substitute
$U_{\left\{ 1,2,3\right\} }$ into (\ref{eq:nsf}) and use the orthogonality
property of the two sets of basis. For $t$-channel we read off the
corresponding vector $V(\Gamma_{t})$ using the graphic rules discussed
in section \ref{sec:amplitude-numerator-formulas}. From Fig. \ref{fig:nhl}
we see that leaf $2$ is connected to leaf $3$ instead of the marked
line, so that $V(\Gamma_{t})=[E_{3},E_{2}]\,V_{1}$. Plugging this
vector into (\ref{eq:nsf}) we see that Jacobi identity is indeed
automatically satisfied.
\begin{equation}
\left\langle U_{\left\{ 1,2,3\right\} },V(\Gamma_{t})\right\rangle =\left\langle U_{\left\{ 1,2,3\right\} },[E_{3},E_{2}]V_{1}\right\rangle =\pi^{2}s_{23}\left(s_{12}-s_{13}\right).
\end{equation}

For higher points, we continue and write down reference vectors using
(\ref{eq:nlsm-ref-vector}), bearing in mind that only cases with
even $n$ correspond to physical scenarios.
\begin{align}
U_{\{1,2,3,4\}}= & 2i\,s_{1234}\pi^{2}\Bigl(\mathcal{S}[234|234](E_{4}E_{3}E_{2}V_{1})^{*}+\mathcal{S}[324|324](E_{4}E_{2}E_{3}V_{1})^{*}\\
 & +\mathcal{S}[423|423](E_{3}E_{2}E_{4}V_{1})^{*}+\mathcal{S}[432|432](E_{2}E_{3}E_{4}V_{1})^{*}+\dots\Bigr).\nonumber 
\end{align}
Consider however acting the above reference vector at $5$-point by
a lowering operator $F_{4}$. Earlier we have derived a recursion
relation (\ref{eq:recuriosn-dual-vectors}) between dual basis vectors
in different weight subspaces. From the recursion relation, we see
that only the two terms in the first line of $U_{\{1,2,3,4\}}$ above
have $E_{4}$ placed at the required position and survive the action
of $F_{4}$. In addition, the $k_{4}$ dependence of these two terms
factorises, so that we may express the result as a lower-point reference
vector.
\begin{align}
F_{4}U_{\{1,2,3,4\}} & =2i\,s_{1234}\pi^{2}(k_{1}+k_{2}+k_{3})\cdot k_{4}\Bigl(\mathcal{S}[23|23](E_{3}E_{2}V_{1})^{*}+\mathcal{S}[32|32](E_{2}E_{3}V_{1})^{*}\Bigr)\\
 & =i(\frac{s_{1234}}{s_{123}}\sum_{i=1,2,3}s_{i4})\,U_{\{123\}}.\nonumber 
\end{align}
The actions of $E_{2}$ and $E_{3}$ are the same as swapping $k_{4}\leftrightarrow k_{2}$
and $k_{3}$ respectively, because of the permutation invariance of
the reference vectors. For generic $n$ we have the recursion relation
between reference vectors.
\begin{equation}
F_{j}U_{\{1,\dots,n-1\}}=\Bigl(i\frac{s_{123\dots n-1}}{s_{\{1,\dots,n-1\}\backslash\{j\}}}\sum_{i\in\left\{ 1,\dots,n-1\right\} \backslash j}s_{ij}\Bigr)U_{\{1,\dots,n-1\}\backslash\{j\}}.\label{eq:urecur}
\end{equation}
It is straightforward to see that conversely all reference vectors,
and therefore all amplitudes and numerators of the nonlinear sigma
model can be determined by this recursion relation and the permutation
symmetry of reference vectors.

\subsection{Structure constant formulation of nonlinear sigma model}

\label{sec:structure-constants}

Seeing that all nonlinear sigma model numerators are expressible as
Shapovalov form in a unified setting, a natural question is then whether
we can take one step further and write the numerators in terms of
products of structure constants, so that the amplitudes can be manifestly
$\phi^{3}$-like. As it turns out, for this purpose we need the notion
of a Shapovalov form defined on the Lie algebra itself, instead of
just the Shapovalov form defined on the Verma module $M_{k_{1}}$,
which is actually a stronger condition. To distinguish the two, let
us denote the presumed kinematic algebra by lowercase letters. The
raising operators in the Chevalley basis are given by $e_{i}$, with
the roots $k_{1}$, $k_{2}$, $\dots$ identified with the particle
momenta, and similarly for the $f_{i}$ and $h_{k_{i}}$. These satisfy
the same algebra relations as before (\ref{eq:chevalley-basis-relations}),
and the Shapovalov form on the Lie algebra $g$ is defined to be the
symmetric bilinear map: $g\times g\rightarrow\mathbb{C}$ that satisfies
the analogue of (\ref{eq:shapovalov-form-defn}).
\begin{equation}
\left\langle [e_{i},v],u\right\rangle =\left\langle v,[f_{i},u]\right\rangle ,\label{eq:algebra-shapovalov-defn}
\end{equation}
for $u$, $v$ that are Lie algebra elements instead of formal vectors.
The above is a stronger requirement because we can readily recover
the Verma module version by manually selecting one of the root vectors,
for example $e_{1}$, to be the lowest weight vector $V_{1}$ and
denoting $E_{i}=ad(e_{i})$ as the adjoint action, but not the other
way around.

The notion of a Shapovalov form on the Lie algebra allows us to define
the norms of root vectors --- the purpose being that a $\phi^{3}$-like
expression only makes sense when the physical information is indeed
introduced through the structure constants instead of entirely through
norms, and for this reason we need constant normalisation so that
all the momentum dependence is carried by the structure constants
and not the opposite. In the standard QCD calculations (see for example
\cite{Dixon:1996wi}) this is achieved by requiring $tr(T^{a}T^{b})=\delta^{a,b}$,
which prevents further rescaling of the commutation relations. In
the following we work with the orthonormal basis vectors $\left\langle e_{(i)},e_{(j)}\right\rangle =\delta_{i,j}$
and we distinguish them from the original root vectors by wrapping
labels with parenthesis.

In the $4$-point example discussed earlier, there was only one root
vector that has root $k_{1}$, $k_{2}$ and $k_{3}$ respectively,
these are related to the orthonormal basis vectors by $e_{(1)}=ie_{1}$,
$e_{(2)}=ie_{2}$ and $e_{(3)}=ie_{3}$. Next, we have non-simple root
vectors $[e_{2},e_{1}]$ and $[e_{3},e_{1}]$, and their norms are
given by the Shapovalov form using the defining property (\ref{eq:algebra-shapovalov-defn}),
\begin{equation}
\left\langle [e_{2},e_{1}],[e_{2},e_{1}]\right\rangle =s_{12},\hspace{0.5cm}\left\langle [e_{2},e_{1}],[e_{3},e_{1}]\right\rangle =s_{13},
\end{equation}
so that the noramlised vectors should be $e_{(4)}=-[e_{(2)},e_{(1)}]/\sqrt{s_{12}}$
and $e_{(5)}=-[e_{(3)},e_{(1)}]/\sqrt{s_{13}}$ respectively. Finally
there are two elements $[e_{3},[e_{2},e_{1}]]$ and $[e_{2},[e_{3},e_{1}]]$
sharing the same root $k_{1}+k_{2}+k_{3}$. Orthonormalising similarly
yields
\begin{align}
e_{(6)} & =\frac{1}{\sqrt{\left(s_{13}+s_{23}\right)s_{12}}}[e_{(3)},[e_{(2)},e_{(1)}]],\\
e_{(7)} & =\frac{i}{\sqrt{s_{13}s_{23}\left(s_{13}+s_{23}\right)\left(s_{12}+s_{13}+s_{23}\right)}}\left(s_{13}[e_{(3)},[e_{(2)},e_{(1)}]]-\left(s_{13}+s_{23}\right)s_{12}[e_{(2)},[e_{(3)},e_{(1)}]]\right).\nonumber 
\end{align}
At $4$-point the above collection $\{e_{(1)},e_{(2)},\dots,e_{(7)}\}$
are the orthonormal basis vectors we need. Having a basis specified,
it is then straightforward to work out the structure constants.
\begin{align}
f_{(1),(2)}{}^{(4)} & =\sqrt{s_{12}},\hspace{0.5cm}f_{(1),(3)}{}^{(5)}=\sqrt{s_{13}},\\
f_{(3),(4)}{}^{(6)} & =-\sqrt{\left(s_{13}+s_{23}\right)},\nonumber \\
f_{(2),(5)}{}^{(6)} & =-\sqrt{\frac{s_{12}s_{13}}{s_{13}+s_{23}}},\hspace{0.5cm}f_{(2),(5)}{}^{(7)}=\frac{\sqrt{s_{23}\left(s_{13}+s_{23}\right)s_{123}}}{s_{13}+s_{23}},\nonumber 
\end{align}
with all others being zeros.

To translate the nonlinear sigma model numerator into structure constants,
we start with the reference vector (\ref{eq:4pt-nlsm-ref-vector}).
In terms of orthonormal basis vectors, this is given by
\begin{align}
U_{\{1,2,3\}} & =-\pi^{2}\left(\frac{s_{12}+s_{23}}{s_{123}}[e_{3},[e_{2},e_{1}]]+\frac{s_{13}+s_{23}}{s_{123}}[e_{2},[e_{3},e_{1}]]\right),\\
 & =\pi^{2}i\sqrt{s_{12}\left(s_{13}+s_{23}\right)}\,e_{(6)}-\pi^{2}i\sqrt{\frac{s_{13}s_{23}\left(s_{13}+s_{23}\right)}{s_{123}}}\,e_{(7)}
\end{align}
so that for example the $s$-channel numerator can be written as
\begin{align}
N(\Gamma_{s}) & =-i\left\langle U_{\{1,2,3\}},[e_{(3)},[e_{(2)},e_{(1)}]]\right\rangle \\
 & =-\pi^{2}\sqrt{s_{12}\left(s_{13}+s_{23}\right)}\,f_{(1),(2)}{}^{(4)}\,f_{(4),(3)}{}^{(6)},\nonumber 
\end{align}
where we translated the graphic rules assigned vector $V(\Gamma_{s})=E_{3}E_{2}V_{1}$
into our current settings, using $E_{i}\rightarrow ad(e_{i})$ and
$V_{1}\rightarrow e_{1}$. The above numerator is identical to the
previously obtained results when explicit values of the structure constants
are plugged in.

\section{Extending to gauge theory}
\label{sec:heft-numerators}
In the next example, we consider the pre-numerators written down explicitly
in the context of heavy-mass effective field theory (HEFT) \cite{Brandhuber:2021kpo,Brandhuber:2021bsf}.
These are the half ladders with the side rail (the line connecting
legs $1$ and $n$) being a massive scalar or a particle with spin,
and the rest of the legs are gluons, minimally coupled to the massive
particle. The presence of a heavy massive line should not be considered
as restriction because the pure Yang-Mills numerators can be fully
recovered once all the pre-numerators are known. This is due to the
insight that in four dimensions the pair of spinors carried by legs
$1$ and $n$ together behave like a polarisation in the spinor-helicity
formulation \cite{Chen:2019ywi,Chen:2021chy}, while
the single gluon insertions behave like the Clifford algebra. The
the observation was proved to generalise to higher dimensions and is of
great practical importance to gravitational wave calculations.
Explicitly the $4$-point pre-numerator reads
\begin{equation}
\mathcal{N}(12,v)=-\,\frac{v\cdot F_{1}\cdot F_{2}\cdot v}{2\,v\cdot k_{1}},\label{eq:4pt-heft-numerator}
\end{equation}
where $v$ denotes the massive-particle velocity and gluon momenta
of legs $2$, $3$, $\dots$, $n-1$ are denoted as $k_{1}$, $k_{2}$,
$\dots$, $k_{n-2}$ respectively. A systematic set of rules for writing
down directly pre-numerators of all multiplicities were given in \cite{Brandhuber:2021bsf},
where all polarisation dependence was introduced through the gauge
invariant stress tensor $F_{i}^{\mu\nu}:=k_{i}^{\mu}\epsilon_{i}^{\nu}-\epsilon_{i}^{\mu}k_{i}^{\nu}$.
Our goal here is to formulate pre-numerators such as equation (\ref{eq:4pt-heft-numerator})
above in algebra language similar to (\ref{eq:nsf}) and to the nonlinear
sigma model example just discussed. 

An immediate problem we face in the presence of gauge theory is how
to incorporate polarisations and gauge degrees of freedom into the
algebra picture. In our present construction roots of the presumed
kinematic algebra is strictly tied to particle momenta, which are
responsible for the algebra explanations of both momentum kernel and
propagator matrix. A solution to this apparent problem can be borrowed
from string theory, in particular when the momenta is restricted
to integer lattice. It was observed by Goddard and Olive \cite{Goddard:1983at}
that the gluon vertex operator in standard bosonic open string theory
can be regarded as the action of a tachyon on another tachyon, followed
by a closed contour integral, provided the scalar product between
the two momenta $q_{1}\cdot q_{2}$ takes particular negative integer
value $-2$. In this case residue theorem translates the integral into
a derivative, yielding
\begin{align}
\frac{1}{2\pi i}\oint dt\,\frac{e^{iq_{2}\cdot X(t)}}{t}\,\frac{e^{iq_{1}\cdot X(z)}}{z} & =\frac{1}{2\pi i}\oint dt\,(t-z)^{q_{2}\cdot q_{1}}:e^{iq_{2}\cdot X(t)}e^{iq_{1}\cdot X(z)}:\label{eq:vertex-algebra}\\
 & =iq_{2}\cdot X'(z)\,e^{i(q_{1}+q_{2})\cdot X(z)},\nonumber 
\end{align}
which is formally identical to a gluon vertex operator if we identify
the polarisation and gluon momentum by $\epsilon=q_{2}$ and $k=q_{1}+q_{2}$
respectively. Massless condition of the gluon $k^{2}=0$ and gauge
invariance $\epsilon\cdot k=0$ are automatically guaranteed by the
mass shell conditions of the two tachyons and $q_{1}\cdot q_{2}=-2$.
Vertex operators of higher excitation modes can be generated similarly
from other choices of the scalar products and repeated actions of
the tachyons. This construction is known to be a representation
of the (generically affine) Lie algebras, with the tachyon actions
being simple root vectors $E_{q}$. 

The spectrum-generating algebra observed by Goddard and Olive in string
theory suggests that we modify our root system in the presence of
gauge theory as the following. For each gluon, we introduce a pair
of simple roots $\{\epsilon_{i},\,\delta_{i}\}$ so that for all $n-2$
gluon legs we need together $2n-4$ roots in the Lie algebra. We may
incorporate all these roots into a unified notation if we like, and
write
\begin{equation}
\alpha_{j}=\begin{cases}
\epsilon_{\left\lfloor j/2\right\rfloor +1} & j\text{ odd},\\
\delta_{j/2} & j\text{ even}.
\end{cases}
\end{equation}
We do not demand all roots to be on the integer lattice, but only
that each pair satisfies the following relations.
\begin{equation}
\epsilon_{i}^{2}=\delta_{i}^{2}=2,\;\epsilon_{i}\cdot\delta_{i}=-2.\label{eq:root-relations}
\end{equation}
The simple roots are normalised to have lengths $2$, which is standard
in classical Lie algebras. The algebra relations are the same as (\ref{eq:chevalley-basis-relations}).
In Chevalley basis we have explicitly,
\begin{equation}
[E_{i},F_{j}]=\delta_{ij}H_{k_{i}},\hspace{0.5cm}[H_{k_{i}},E_{j}]=(\alpha_{i}\cdot\alpha_{j})E_{j},\hspace{0.5cm}[H_{k_{i}},F_{j}]=-(\alpha_{i}\cdot\alpha_{j})F_{j},
\end{equation}
for $i$, $j=1$, $2$, $\dots$, $2n-4$. Following the same spirit
as (\ref{eq:vertex-algebra}), each gluon in this picture is represented
algebraically by the commutator 
\begin{equation}
E_{k_{i}}:=[E_{\epsilon_{i}},E_{\delta_{i}}],\label{eq:non-simple-root-vect}
\end{equation}
with root $k_{i}:=\delta_{i}+\epsilon_{i}$, so that it is still a
root vector of the Lie algebra, albeit a non-simple one. We define
the Verma module $M_{v}$ accordingly as the vector space generated
by $V_{v}$ with its weight equal to the velocity $v$ of the massive-particle.
In this section we focus only on the pre-numerators, assuming that
amplitudes can be readily recovered via double copies. For simplicity
we rescale the reference vector in (\ref{eq:nsf}) to absorb the overall
factor $1/2\,s_{123\dots n-1}$. The pre-numerators are then given
by Shapovalov form on $M_{v}$ in the natural way.
\begin{align}
{\cal N}(1\dots n-2,v) & =\left\langle U(1\dots n-2,v),E_{k_{n-2}}\dots E_{k_{1}}V_{v}\right\rangle \label{eq:heft-numerator-shapovalof-form}\\
 & =\left\langle U(1\dots n-2,v),[E_{\epsilon_{n-2}},E_{\delta_{n-2}}]\dots[E_{\epsilon_{2}},E_{\delta_{2}}]\,[E_{\epsilon_{1}},E_{\delta_{1}}]V_{v}\right\rangle ,\nonumber 
\end{align}
where we used relation (\ref{eq:non-simple-root-vect}) to write the
right hand side as a linear combination of the basis vectors $V(\sum_{i=1}^{n-2}k_{i},\sigma)=E_{\alpha_{\sigma(sn-4)}}\dots E_{\alpha_{\sigma(2)}}E_{\alpha_{\sigma(1)}}V_{v}$.

One consequence of interpreting gauge theory particles as composite
algebraic object is that the reference vector $U(1\dots n-2,v)$ appears
in the Shapovalov form formula (\ref{eq:heft-numerator-shapovalof-form})
is no longer unique (compared to the previous nonlinear sigma model
example). This is because for the formula (\ref{eq:heft-numerator-shapovalof-form})
to produce the correct pre-numerators we only need the Shapovalov
forms of $U(1\dots n-2,v)$ evaluated together with the $(n-2)!$
vectors corresponding to half ladders. The collection of all such
vectors only form a subspace of the full Verma module because the
root vectors $E_{\epsilon_{i}}$ and $E_{\delta_{i}}$ introduced
by each gluon are always adjacent. One way to streamline this algebraic
setting is to work instead with the module quotient by the position
of one root vector, for example $E_{\epsilon_{i}}$, relative to the
other, so that we have a generic $g$-module instead of just the free
module. Whether the introduction of such an equivalence class is a necessary
operation that reflexes gauge degrees of freedom algebraically is
still very much an open question to the authors. In the following
we will continue with the Verma module language without any quotient
operation.

We would like to emphasise that the above algebraic construct is based
on purely field theory settings, even though motivations were found
in string theory. Note however that in the Shapovalov form formulation,
the pre-numerators (\ref{eq:heft-numerator-shapovalof-form}) bear
a close resemblance to their string theory counterparts described in
\cite{Fu:2018hpu}. From the algebra perspective, the numerators
of field theories and string are formally closer, while the notion
of field theory amplitudes naturally derives from the Shapovalov form
on the dual basis (\ref{eq:asf}). With that being said, one may regard
the introduction of the Shapovalov form as a convenient ansatz where numerators
are spanned by momentum kernels, with half of the momenta substituted
by polarisations, if one would prefer to proceed from a purely field
theory perspective. In particular, it would be interesting to see if
the ansatz approach can be generalised to loop levels.

At $4$-point, one choice for the reference vector that fits our purpose
is the following.
\begin{align}
U(12,v) & =\frac{\epsilon_{2}\cdot v}{4p_{1}\cdot v}\left\langle [E_{\text{\ensuremath{\delta}}_{1}},[E_{\epsilon_{1}},[E_{\delta_{2}},E_{\epsilon_{2}}]]]V_{v},\,E_{\epsilon_{2}}E_{\delta_{2}}E_{\epsilon_{1}}E_{\delta_{1}}V_{v}\right\rangle (E_{\epsilon_{2}}E_{\delta_{2}}E_{\epsilon_{1}}E_{\delta_{1}}V_{v})^{*}\label{eq:4pt-ref-vector}\\
 & +(\epsilon_{1}\leftrightarrow\delta_{1})+(\epsilon_{2}\leftrightarrow\delta_{2})+(\epsilon_{1}\leftrightarrow\delta_{1},\epsilon_{2}\leftrightarrow\delta_{2})\nonumber \\
 & +(1\leftrightarrow2)\nonumber 
\end{align}
The second line above is obtained from the first line with $\epsilon_{1}\leftrightarrow\delta_{1}$
swapped, or $\epsilon_{2}\leftrightarrow\delta_{2}$, or both swapped,
while the third line denotes terms obtained by swapping the labels
$1\leftrightarrow2$ from everything in the first two lines of the
equations (but with symbols such as $\epsilon$'s and $\delta$'s
unchanged).

To see how the reference vector above produces exactly (\ref{eq:4pt-heft-numerator})
when plugged into the Shapovalov form (\ref{eq:heft-numerator-shapovalof-form}),
first note the following algebra identity between nested commutators
(which will soon become useful as well when explaining the reference
vectors at higher points).
\begin{equation}
F_{i}[[[A,E_{i}],B]\dots C]\,V_{v}=\alpha_{i}\cdot\alpha_{A}[[A,B],\dots C]\,V_{v},\label{eq:nested-commutator-identity}
\end{equation}
where $A$, $B$ and $C$ denote polynomials\footnote{One needs to modify (\ref{eq:nested-commutator-identity}) to be a
linear sum, if $A$ is a linear combination of terms of different
weight.} of raising operators $E_{j}$, or may be nested commutators themselves,
as long as they do not contain any factor of $E_{i}$. The identity
(\ref{eq:nested-commutator-identity}) applies to an arbitrary number
of nested commutators. The fact that this identity is true can be
verified by moving the lowering operator $F_{i}$ to the right, seeing
that $A$, $B$ and $C$ do not contain $E_{i}$ and therefore commute
with the lowering operator. The $F_{i}$ eventually reaches the lowest
weight vector $V_{1}$ and annihilates, leaving only the commutator
$[F_{i},E_{i}]=-H_{\alpha_{i}}$ that replaces the original $E_{i}$
at its position in (\ref{eq:nested-commutator-identity}). As a
result, we need to consider the commutator of this Cartan subalgebra
element, $-[[[A,H_{\alpha_{i}}],B]\dots C]\,V_{v}$. The action of
$H_{\alpha_{i}}$ yields all the roots on its right and the lowest
weight combined, so that the commutator yields a factor $\alpha_{i}\cdot\alpha_{A}$
as was claimed by the identity.

Additionally, note the polarisation dependence on the expected outcome
(\ref{eq:4pt-heft-numerator}) is introduced entirely through stress
tensors, which are anti-symmetric and therefore invariant under the
shifting $k_{i}\rightarrow\delta_{i}=k_{i}-\epsilon_{i}$. That is,
$\delta_{i}^{\mu}\epsilon_{i}^{\nu}-\epsilon_{i}^{\mu}\delta_{i}^{\nu}=k_{i}^{\mu}\epsilon_{i}^{\nu}-\epsilon_{i}^{\mu}k_{i}^{\nu}=F_{i}$.
Bearing this in mind, we see that the pre-numerator is a product of
$v$ and $\delta_{i}$'s and $\epsilon_{j}$'s, plus terms with various
$\epsilon_{i}\leftrightarrow\delta_{i}$ swaps.
\begin{align}
-\,\frac{v\cdot F_{1}\cdot F_{2}\cdot v}{2\,v\cdot k_{1}} & =\frac{-1}{2\,v\cdot k_{1}}(v\cdot\delta_{1})(\epsilon_{1}\cdot\delta_{2})(\epsilon_{2}\cdot v)\label{eq:4pt-symmetrisation}\\
 & -(\epsilon_{1}\leftrightarrow\delta_{1})-(\epsilon_{2}\leftrightarrow\delta_{2})+(\epsilon_{1}\leftrightarrow\delta_{1},\epsilon_{2}\leftrightarrow\delta_{2}).\nonumber 
\end{align}
As a consequence, for the purpose of verifying that the $4$-point
reference vector given by (\ref{eq:4pt-ref-vector}) indeed produces
the expected formula (\ref{eq:4pt-heft-numerator}), we only need
to check that the Shapovalov form contains one such product term and
satisfies the required symmetry, and we proceed by taking the Shapovalov
form of reference vector with all basis vectors.

Let us begin with the Shapovalov form of $[E_{\text{\ensuremath{\delta}}_{1}},[E_{\epsilon_{1}},[E_{\delta_{2}},E_{\epsilon_{2}}]]]V_{v}$,
which is simpler, instead of the Shapovalov form of the full $U(12,v)$.
For generic momentum configuration (aside from the conditions imposed
by (\ref{eq:root-relations})) the other slot of the Shapovalov form
must be $V_{1}$ raised by the same set of operators. Let us begin
with just one of such vectors, $E_{\epsilon_{2}}E_{\delta_{2}}E_{\epsilon_{1}}E_{\text{\ensuremath{\delta}}_{1}}V_{1}$.
We move $E_{\epsilon_{2}}$ from the right slot to the left, using
the defining property of Shapovalov form (\ref{eq:shapovalov-form-defn}),
translating the raising operator into $F_{\epsilon_{2}}$. The action
of this operator is in turn given by the identity just proved (\ref{eq:nested-commutator-identity}),
which in this case produces a factor $\epsilon_{2}\cdot\delta_{2}=-2$,
as it is required by our root system (\ref{eq:root-relations}).
\begin{align}
\left\langle [E_{\text{\ensuremath{\delta}}_{1}},[E_{\epsilon_{1}},[E_{\delta_{2}},E_{\epsilon_{2}}]]]V_{v},E_{\epsilon_{2}}E_{\delta_{2}}E_{\epsilon_{1}}E_{\text{\ensuremath{\delta}}_{1}}V_{1}\right\rangle  & =\left\langle F_{\epsilon_{2}}\,[E_{\text{\ensuremath{\delta}}_{1}},[E_{\epsilon_{1}},[E_{\delta_{2}},E_{\epsilon_{2}}]]]V_{v},E_{\delta_{2}}E_{\epsilon_{1}}E_{\text{\ensuremath{\delta}}_{1}}V_{1}\right\rangle \\
 & =\epsilon_{2}\cdot\delta_{2}\left\langle F_{\epsilon_{2}}\,[E_{\text{\ensuremath{\delta}}_{1}},[E_{\epsilon_{1}},E_{\delta_{2}}]]V_{v},E_{\delta_{2}}E_{\epsilon_{1}}E_{\text{\ensuremath{\delta}}_{1}}V_{1}\right\rangle .\nonumber 
\end{align}
Continuing moving the rest of the raising operators one by one, we arrive
at the following.
\begin{equation}
(\epsilon_{2}\cdot\delta_{2})(\delta_{2}\cdot\epsilon_{1})(\epsilon_{1}\cdot\delta_{1})(\delta_{1}\cdot v)=(-2)^{2}(\delta_{2}\cdot\epsilon_{1})(\delta_{1}\cdot v).\label{eq:4pt-starting-term}
\end{equation}
The above agrees with the first line of equation (\ref{eq:4pt-symmetrisation}),
except for a factor $(\epsilon_{2}\cdot v)$ missing. Producing such
a factor purely from algebra apparently requires the introduction
of an additional root vector $E_{v}$ carrying the root that is identical
to the lowest weight $v$, which complicates the root system further
than it already is (even though this is in principle straightforward).
For this reason, we choose instead to include this factor manually
as a coefficient of the reference vector.

As we explained earlier the resulting vector $-(\epsilon_{2}\cdot v)[E_{\text{\ensuremath{\delta}}_{1}},[E_{\epsilon_{1}},[E_{\delta_{2}},E_{\epsilon_{2}}]]]V_{v}/2\,v\cdot k_{1}$
is not yet the full answer. To fully recover the pre-numerator formula,
which is written in terms of stress tensors, we need to include terms
derived from various swaps. This is not however derived from simply
swapping the vector just obtained. The reason being that (\ref{eq:4pt-starting-term})
was calculated from the Shapovalov form of this vector together with
a specific basis vector $E_{\epsilon_{2}}E_{\delta_{2}}E_{\epsilon_{1}}E_{\text{\ensuremath{\delta}}_{1}}V_{1}$,
and that swapping the $\delta_{i}$'s and $\epsilon_{j}$'s in equation
(\ref{eq:4pt-starting-term}) requires that we swap both slots
of the Shapovalov form. One solution to this issue is that we project
the vector just obtained onto the dual basis vector, and then swap.
The result is the first two lines shown in equation (\ref{eq:4pt-symmetrisation}).
In addition note from the discussions at the end of section \ref{sec:amplitude-numerator-formulas}
that the reference vector needs to be permutation invariant with respect
to the $n-2$ half ladder labels. In our case this corresponds to
permutations with respect to $\{1,2,\dots,n-2\}$ while each pair
$\{\epsilon_{i},\,\delta_{i}\}$ is kept fixed relatively, henceforth
the third line of equation (\ref{eq:4pt-symmetrisation}) above.

When we go to higher points, the pre-numerators are given by the fusion
rules explained in \cite{Brandhuber:2021bsf}. Generically the $n-2$
gluon labels are distributed into various numbers of sets that are
used as labels of the fusion rules generators $T_{(i,\,\dots),\,(j,\,\dots),\,\dots}$.
Each set in turn contributes an ordered product of stress tensors
of the corresponding gluon legs. These contributed factors are then
welded together by $V_{A}^{\mu\nu}=v^{\mu}\sum_{j\in A}k_{j}^{\nu}$
for some subset of the labels $A$. For example,
\begin{equation}
\left\langle T_{\left(1458\right),\left(26\right),\left(37\right)}\right\rangle =\frac{v\cdot F_{1458}\cdot V_{1}\cdot F_{26}\cdot V_{12}\cdot F_{37}\cdot v}{8(v\cdot p_{1})(v\cdot p_{1458})(v\cdot p_{124568})}.\label{eq:9pt-prenumerator}
\end{equation}
The subset $A$ associated with the $V_{A}$ that welds a product
of stress tensors from its left is given by all the label numbers
that are both smaller and appear on the left-hand side of the stress
tensors. Also, there are appropriate scalar products in the denominator.
(See \cite{Brandhuber:2021bsf} and the references therein for more
details and physical origin of the fusion rules.) Our claim is that
any term such as (\ref{eq:9pt-prenumerator}) can be derived from
the vector
\begin{equation}
X_{(1458),(26),(37)}=\frac{(v\cdot\epsilon_{8})(v\cdot\epsilon_{6})(v\cdot\epsilon_{7})\,[[r(1458),r(26)],r(37)]}{(-2)^{8}(v\cdot p_{1})(v\cdot p_{1458})(v\cdot p_{124568})},
\end{equation}
where $r(1458)$, $r(26)$ and $r(37)$ are nested commutators of
the corresponding root vectors.
\begin{align}
r(1458) & =[E_{\delta_{1}},[E_{\epsilon_{1}},[E_{\delta_{4}},[E_{\epsilon_{4}},[E_{\delta_{5}},[E_{\epsilon_{5}},[E_{\delta_{8}},E_{\epsilon_{8}}]]]]]]],\\
r(26) & =[E_{\delta_{2}},[E_{\epsilon_{2}},[E_{\delta_{6}},E_{\epsilon_{6}}]]],\nonumber \\
r(37) & =[E_{\delta_{3}},[E_{\epsilon_{3}},[E_{\delta_{7}},E_{\epsilon_{7}}]]].\nonumber 
\end{align}
The way that the vector $X_{(1458),(26),(37)}$ contributes (\ref{eq:9pt-prenumerator})
is through projection onto the dual basis vector, and then the $\delta_{i}$'s
and $\epsilon_{j}$'s swapped, and gluon labels permuted as in the
$4$-point example just discussed. The full reference vector is given
by the sum of all such vectors, each contributes one term given by
the fusion rules.

The above claim can be verified in a rather straightforward manner.
We take the Shapovalov form of $X_{(1458),(26),(37)}$ with a basis
vector and move raising operators one by one, producing the following
product.
\begin{align}
 & \left\langle [[r(1458),r(26)],r(37)],\,E_{\epsilon_{8}}E_{\delta_{8}}E_{\epsilon_{7}}E_{\delta_{7}}\dots E_{\epsilon_{2}}E_{\delta_{2}}E_{\epsilon_{1}}E_{\text{\ensuremath{\delta}}_{1}}V_{1}\right\rangle \label{eq:9pt-product-1}\\
 & =(-2)^{8}(\delta_{8}\cdot\epsilon_{5})(\delta_{7}\cdot\epsilon_{3})(\delta_{6}\cdot\epsilon_{2})(\delta_{5}\cdot\epsilon_{4})(\delta_{4}\cdot\epsilon_{1})(\delta_{3}\cdot(k_{1}+k_{2}))\dots\label{eq:9pt-product-2}\\
 & \sim\left(v\cdot(\delta_{1}\epsilon_{1})\cdot\dots\cdot\delta_{8}\right)\left(k_{1}\cdot(\delta_{2}\epsilon_{2})\cdot\delta_{6}\right)\left((k_{1}+k_{2})\cdot(\delta_{3}\epsilon_{3})\cdot\delta_{7}\right).\nonumber 
\end{align}
Supplying the product manually with the missing factors $(v\cdot\epsilon_{8})(v\cdot\epsilon_{6})(v\cdot\epsilon_{7})/(-2)^{8}(v\cdot p_{1})(v\cdot p_{1458})(v\cdot p_{124568})$
and we see the result gives the exactly equation (\ref{eq:9pt-prenumerator})
after anti-symmetrising with respect to all pairs $\{\epsilon_{i},\,\delta_{i}\}$.
Note a lowering operator $F_{j}$ commutes with all $E_{i}$ that
carry distinct roots, so that the product was produced following a
similar process at $4$-point, except when acting on the last root
vectors in $r(26)$ and $r(37)$. The resulting Cartan subalgebra
elements do not directly act on $V_{v}$ because they are still inside
the commutators. Instead, a factor of $(\delta_{2}\cdot k_{1})$ and
a $(\delta_{2}\cdot(k_{1}+k_{2}))$ were produced from the remaining
root vectors within the commutator because of the identity (\ref{eq:nested-commutator-identity}).
No roots with labels greater than $2$ and $3$ should contribute to
these factors respectively, because they were eliminated by the lowering
operator $F_{j}$'s prior to the application of identity (\ref{eq:nested-commutator-identity}).
This completes the proof for generic terms generated by fusion rules.

\section*{Acknowledgements} 
We thank Vladimir Dobrev and the organisers of LT-14 for the opportunity to present our work. 
We thank Pierre Vanhove for suggesting the discussion on HEFT numerators (section \ref{sec:heft-numerators}).
CF is grateful for Chao-Qiang Geng and Hangzhou Institute for Advanced Study for their hospitality.
YW is supported by China National Natural Science Funds for Distinguished Young Scholar (Grant No. 12105062) and Agence Nationale de la Recherche (ANR), project ANR-22-CE31-0017

\end{document}